\documentclass[prl,aps,preprint]{revtex4-1}
\usepackage{amsmath}
\usepackage{amssymb}
\usepackage{graphicx}

\newcommand{\mP}{{\mathcal P}}
\newcommand{\mT}{{\mathcal T}}

\newcommand{\beqa}{\begin{eqnarray}}
\newcommand{\eeqa}{\end{eqnarray}}

\begin{document}
\title{Degrees and signatures of broken $\mP\mT$-symmetry in (non-uniform) lattices} 
\author{Derek D. Scott}
\author{Yogesh N. Joglekar}
\affiliation{Department of Physics, 
Indiana University Purdue University Indianapolis (IUPUI), 
Indianapolis, Indiana 46202, USA}
\date{\today}
\begin{abstract}
We investigate the robustness of parity- and time-reversal ($\mP\mT$) symmetric phase in an $N$-site lattice with position-dependent, parity-symmetric hopping function and a pair of imaginary, $\mP\mT$-symmetric impurities. We find that the "fragile" $\mP\mT$-symmetric phase in these lattices is stronger than its counterpart in a lattice with constant hopping.  With an open system in mind, we explore the degrees of broken $\mP\mT$-symmetry and their signatures in single-particle wavepacket evolution. We predict that when the $\mP\mT$-symmetric impurities are closest to each other, the time evolution of a wavepacket in an even-$N$ lattice is remarkably different from that in an odd-$N$ lattice. Our results suggest that $\mP\mT$-symmetry breaking in such lattices is accompanied by rich, hitherto unanticipated, phenomena. 
\end{abstract}
\maketitle

\noindent{\it Introduction:} Since their discovery by Bender and coworkers over a decade ago~\cite{bender1,bender2}, non-Hermitian Hamiltonians $H_{PT}$ that are symmetric with respect to combined parity ($\mP$) and time-reversal ($\mT$) operations have been extensively explored~\cite{mostafa,znojil}. Such continuum Hamiltonians usually feature a non-Hermitian, $\mP\mT$-symmetric potential, $V(x)=V^{*}(-x)\neq V^*(x)$. Although not Hermitian, $H_{PT}\neq H_{PT}^\dagger$, they have purely real eigenvalues over a range of parameters. Traditionally, the region of parameters where the spectrum of the Hamiltonian is real and, therefore, its eigenfunctions are simultaneous eigenfunctions of the combined $\mP\mT$-operation, is defined as the $\mP\mT$-symmetric region. When an eigenvalue of the $\mP\mT$-symmetric Hamiltonian becomes complex, $E\neq E^{*}$, since time-reversal is an anti-linear operation, it follows that the corresponding eigenfunction is not invariant under the $\mP\mT$-operation, $[H_{PT},\mP\mT]\psi_E(x)\neq 0$. Therefore, the number of eigenfunctions that break the $\mP\mT$-symmetry is  equal to the number of complex eigenvalues.  

In the past few years, $\mP\mT$-symmetric lattice models have become a focal point of research~\cite{znojil,bendix,spin,song}. Lattice models, in general, are popular in physics due to their versatility~\cite{wen}, availability of exact solutions~\cite{onsager}, the absence of an ultraviolet divergence~\cite{kogut}, and the ability to capture counterintuitive phenomena that have no counterparts in the continuum theories~\cite{winkler}. Recently, evanescently coupled waveguides~\cite{christo} have emerged as a promising candidate for the realization of an ideal, one-dimensional lattice with tunable hopping~\cite{gf}, disorder~\cite{berg1}, and non-Hermitian, on-site, impurity potentials~\cite{makris,expt1,expt2}. The hopping between adjacent waveguides is controlled by the width of the barrier between the two, and the complex, on-site potential is proportional to the local refractive index, the imaginary part of which provides the loss or gain in the respective waveguide. The emergence of complex eigenvalues in such a pair of waveguides, via the attendant non-reciprocal behavior of light intensity, has been experimentally explored~\cite{expt1,expt2}. A salient feature of a $\mP\mT$-symmetric lattice with (approximately) constant hopping is that its $\mP\mT$-symmetric phase is fragile except when the non-Hermitian, $\mP\mT$-symmetric impurities are closest or the farthest~\cite{bendix,song,mark}. The $\mP\mT$-symmetric phase becomes robust if, instead, a non-Hermitian, $\mP\mT$-symmetric hopping function is chosen~\cite{ya}. However, the degrees of $\mP\mT$-symmetry breaking, defined as the fraction of eigenvalues that become complex, and their consequences to the properties of the system, have not been explored~\cite{henning}. 

In this paper, we investigate the robustness of $\mP\mT$-symmetric phase, and the degrees of $\mP\mT$-symmetry breaking in an $N$-site lattice with a hopping function $t_\alpha(k)=t_0[k(N-k)]^{\alpha/2}$ and a pair of $\mP\mT$-symmetric impurities $\pm i\gamma$ at positions $(m_0,\bar{m}_0)$ respectively where $\bar{m}_0=N+1-m_0$ is the mirror-symmetric counterpart of site $m_0$. We define the $\mP\mT$-symmetric phase as robust (fragile) provided the critical impurity strength $\gamma_{PT}(N)$, below which all eigenvalues are real, is nonzero (zero) in the limit $N\rightarrow\infty$. With coupled optical waveguides in mind~\cite{expt1,expt2}, we explore the signatures of $\mP\mT$-symmetry breaking in the evolution of single-particle properties across the $\mP\mT$-symmetric phase boundary. 

Our main results are as follows: i) the $\mP\mT$-symmetric phase is robust when $\alpha>0$ and the impurities are closest to each other, ii) in the fragile case, the scaling of $\gamma_{PT}(N)$ depends on the fractional impurity position $\mu=m_0/N$ and $\alpha$, iii)  in the robust case, single-particle wavepacket evolution across the $\mP\mT$-symmetric phase boundary is qualitatively different for even and odd $N$. Our results show that, based on the position of impurities and the form of the hopping function, signatures of the $\mP\mT$-symmetry breaking in lattices with $N\gg 1$ are dramatically different from those in the experimentally explored $N=2$ case~\cite{expt1,expt2}. 

\noindent{\it Tight-binding Model:} We start with the Hamiltonian for a one-dimensional lattice 
\begin{equation}
\label{eq:h}
\hat{H}_\alpha=-\sum_{i=1}^{N-1} t_\alpha(i)\left(a^{\dagger}_{i+1} a_i + a^{\dagger}_i a_{i+1}\right) + i\gamma\left(a^{\dagger}_{m_0}a_{m_0}-a^{\dagger}_{\bar{m}_0}a_{\bar{m}_0}\right), 
\end{equation}
where $a^{\dagger}_k (a_k)$ is the creation (annihilation) operator for a state localized at site $k$, and $1\leq m_0\leq N/2$ is the position of the first impurity. We recall that for a constant hopping model, as $\gamma/t_0$ is increased, the maximum number of complex eigenvalues is given by $2m_0=2\mu N\leq N$. This process is always sequential: the number of complex eigenvalues increases by two or four as $\gamma$ increases; the sole exception is the case with nearest neighbor impurities, $m_0=N/2$, where all eigenvalues simultaneoulsy become complex across the $\mP\mT$-symmetric threshold~\cite{mark}. In the following, we obtain $\alpha$-dependent generalizations of these results.  When $\alpha>0$, the bandwidth of a clean lattice scales as $\Delta'_\alpha(N)\sim N^\alpha$; to be consistent with the $\alpha=0$ case, we use $\gamma/\Delta_\alpha$ as the dimensionless measure of the impurity strength where $\Delta_\alpha=\Delta'_\alpha/4$.

\begin{figure}[htb]
\begin{center}
\begin{minipage}{20cm}
\begin{minipage}{9cm}
\hspace{-4cm}
\includegraphics[angle=0,width=8.5cm]{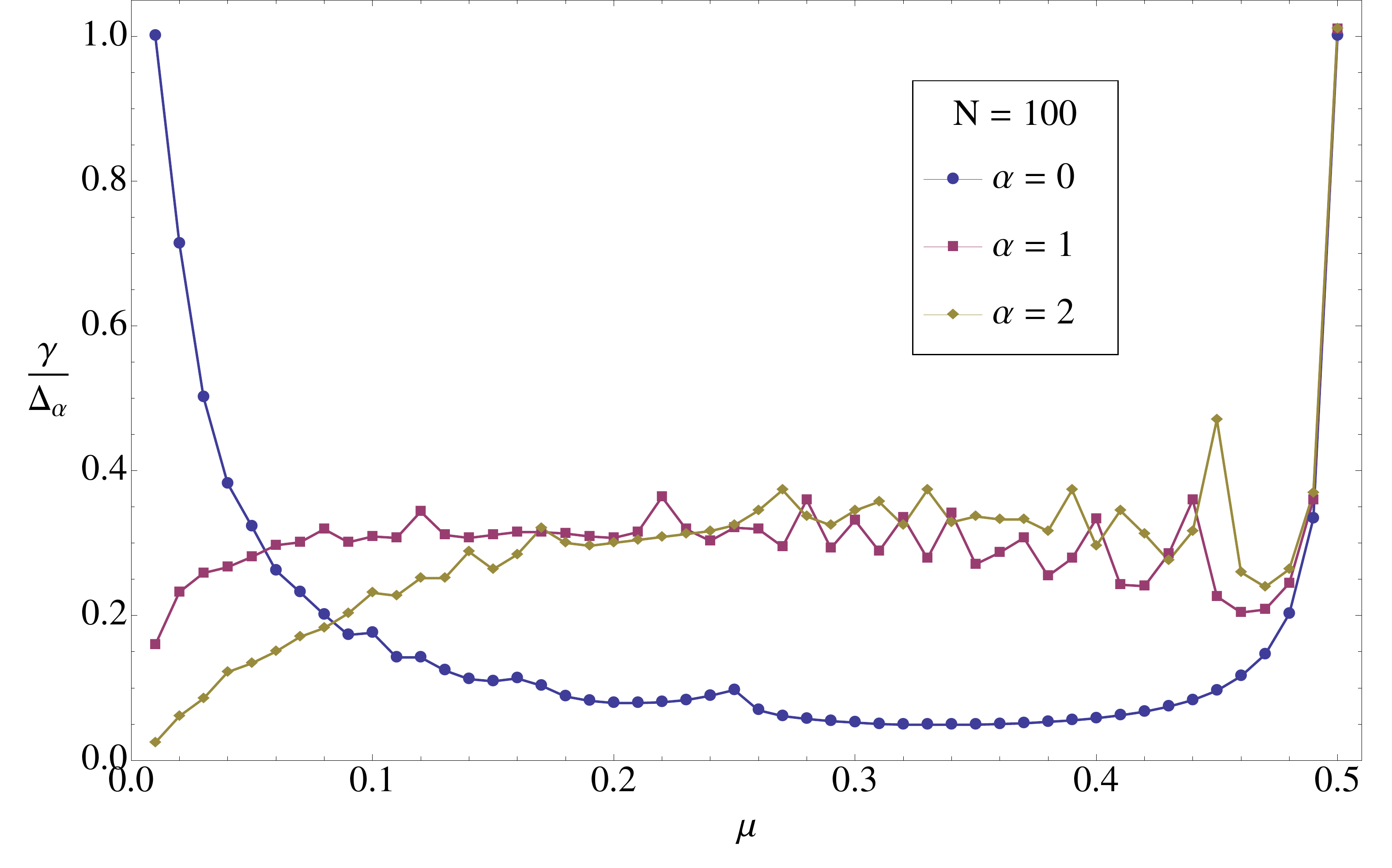}
\end{minipage}
\begin{minipage}{9cm}
\hspace{-4.5cm}
\includegraphics[angle=0,width=9cm]{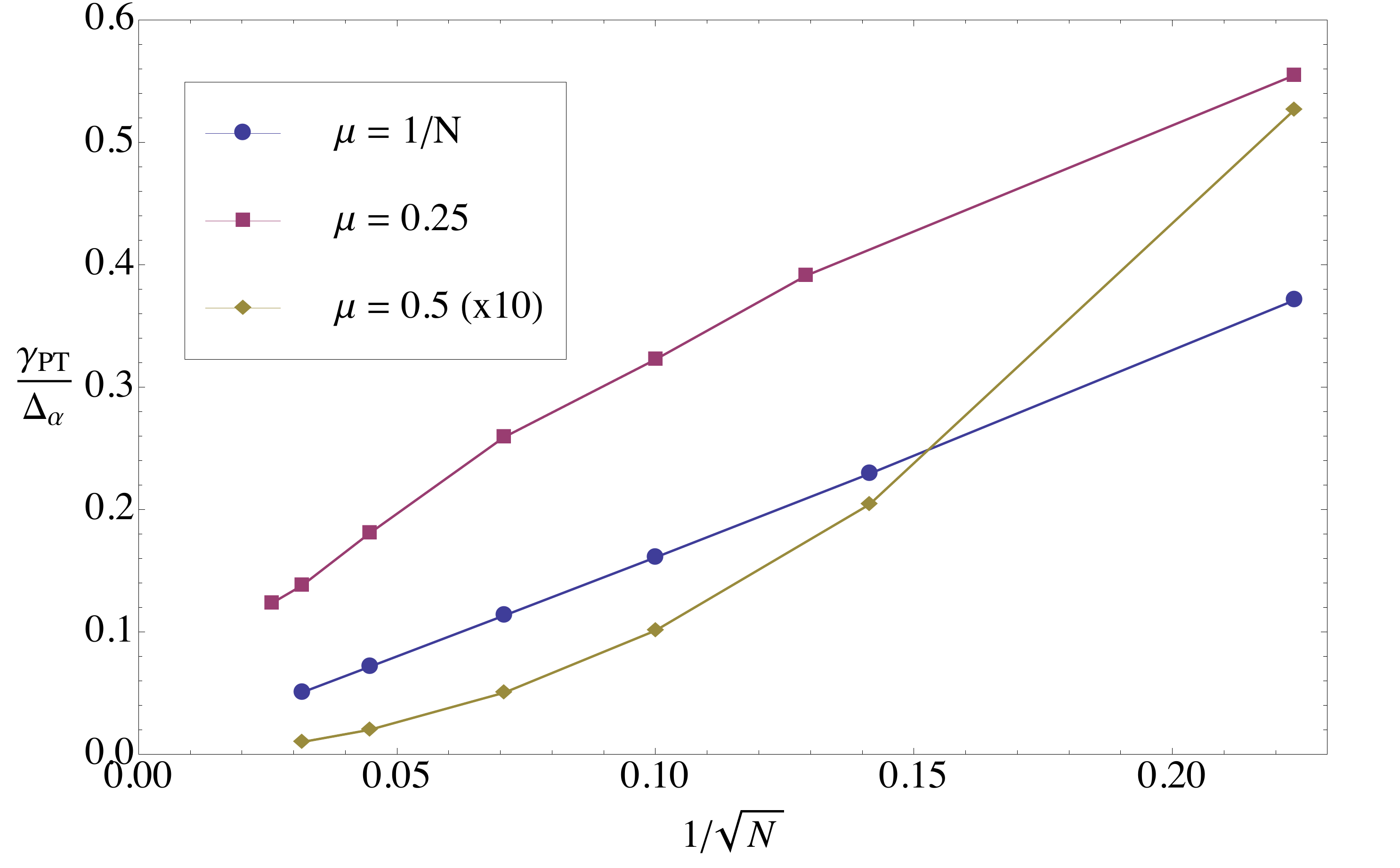}
\end{minipage}
\end{minipage}
\caption{(color online) a) Typical phase diagram of the Hamiltonian $H_\alpha$ as a function of impurity strength $\gamma/\Delta_\alpha$ and fractional impurity position $\mu=m_0/N$ for an $N=100$ site lattice. The region in the $(\gamma/\Delta_\alpha,\mu)$ plane below each curve represents the $\mP\mT$-symmetric phase with purely real eigenvalues; recall that the quarter-bandwidth $\Delta_\alpha\sim N^\alpha$. The $\mP\mT$-symmetric phase is robust at $\mu=1/2$. The results are similar for an odd $N$, except that the critical strength when the impurities are closest is $\gamma_{PT}=\Delta_\alpha/2$. b) Scaling of the dimensionless critical impurity strength $\gamma_{PT}(N,\mu)/\Delta_\alpha$ for an $\alpha=1$ lattice. When the impurities are farthest, $\gamma_{PT}\sim 1/\sqrt{N}$ (blue circles). For an intermediate position, $\mu=1/4$, the critical $\gamma_{PT}\sim N^{-1/3}$ shows sub-linear behavior when plotted as a function of $1/\sqrt{N}$ (red squares). For $\mu=1/2$, the approach to the asymptotic limit $\gamma_{PT}(N)-\gamma_{PT}(N\rightarrow\infty)\sim 1/N$, plotted here, shows a quadratic behavior as a function of $1/\sqrt{N}$ (beige diamonds). Thus, the "fragile" phase of an $\alpha>0$ lattice is significantly stronger than its $\alpha=0$ counterpart.} 
\label{fig:phase}
\end{center}
\end{figure} 
The left panel in Fig.~\ref{fig:phase} shows the typical phase diagram for an even lattice ($N=100$) in the $(\gamma/\Delta_\alpha,\mu)$ plane as a function of the hopping exponent $\alpha\geq 0$~\cite{note}. The $\mP\mT$-symmetric threshold, below which all eigenvalues are real, decreases with increasing distance between the impurities, $d=1+N(1-2\mu)$, but approaches a universal critical value $\gamma_{PT}/\Delta_\alpha=1$ when the impurities are the closest, $\mu=1/2$.  When $N$ is odd (not shown), the fragile nature of the $\mP\mT$-symmetric phase is maintained, but the critical strength for closest impurities is reduced by a factor of two, $\gamma_{PT}=\Delta_\alpha/2$. In both cases, when $\alpha>0$, the critical impurity strength $\gamma_{PT}(\mu)$ for a generic $\mu$ is appreciable compared to its respective universal value at $\mu\approx 1/2$. Thus, the robust nature of the $\mP\mT$-symmetric phase at $\mu=1/N$ and its extremely fragile nature at generic values of $\mu$ ($\alpha=0$, blue circles) are anomalous features of a lattice with constant hopping~\cite{mark}.  

The right panel in Fig.~\ref{fig:phase} shows typical scaling of the critical impurity strength $\gamma_{PT}(N,\mu)$ for $\mu=\{1/N,1/4,1/2\}$ in a lattice with $\alpha=1$. When the impurities are farthest, $\mu=1/N$, the dimensionless critical impurity strength vanishes as $\gamma_{PT}(N)\sim N^{-1/2}$ (blue circles); when $\mu=1/4$, it vanishes as $\gamma_{PT}(N)\sim N^{-1/3}$ (red squares); and when $\mu=1/2$, it approaches the nonzero critical value as $\gamma_{PT}(N)-\gamma_{PT}(N\rightarrow\infty)\sim N^{-1}$ (beige diamonds). Note that these results are in stark contrast with the $\alpha=0$ case where $\gamma_{PT}(N)\sim N^{-1}$ for all generic $\mu$~\cite{mark}. Thus, although the $\mP\mT$-symmetric phase of a non-uniform lattice for a generic impurity position is, in principle, algebraically fragile, {\it the critical impurity strength is still appreciably large for experimentally relevant numbers} (of waveguides); $\gamma_{PT}(N)/\Delta_\alpha\sim 0.5-0.3$ for $N\lesssim 100$.

Next, we consider the emergence of complex eigenvalues~\cite{henning}. For a tight-binding Hamiltonian with a purely imaginary potential, if $E$ is an eigenvalue, then so are $-E$ and $E^{*}$~\cite{mapping}. For a given impurity position $m_0$, the first (four) complex eigenvalues emerge as two adjacent energy levels $(-E,-E+\delta)$ become degenerate, and so do their positive counterparts. Thus, as the fractional impurity position $\mu$ changes from the edge of the lattice ($\mu=1/N$) to its center ($\mu\approx 1/2$), the position of the first complex eigenvalue changes from the center of the energy band, $E=0$, to the its edges, $E\sim\pm 2\Delta_\alpha$. For a given $\mu$, when the impurity strength $\gamma/\Delta_\alpha$ increases, we find that the maximum number of complex eigenvalues is still given by $2\mu N=2m_0\leq N$, and that {\it the process is sequential except in an even lattice at} $\mu=1/2$~\cite{mark}. 

\begin{figure}[htb]
\begin{center}
\includegraphics[angle=0,width=9cm]{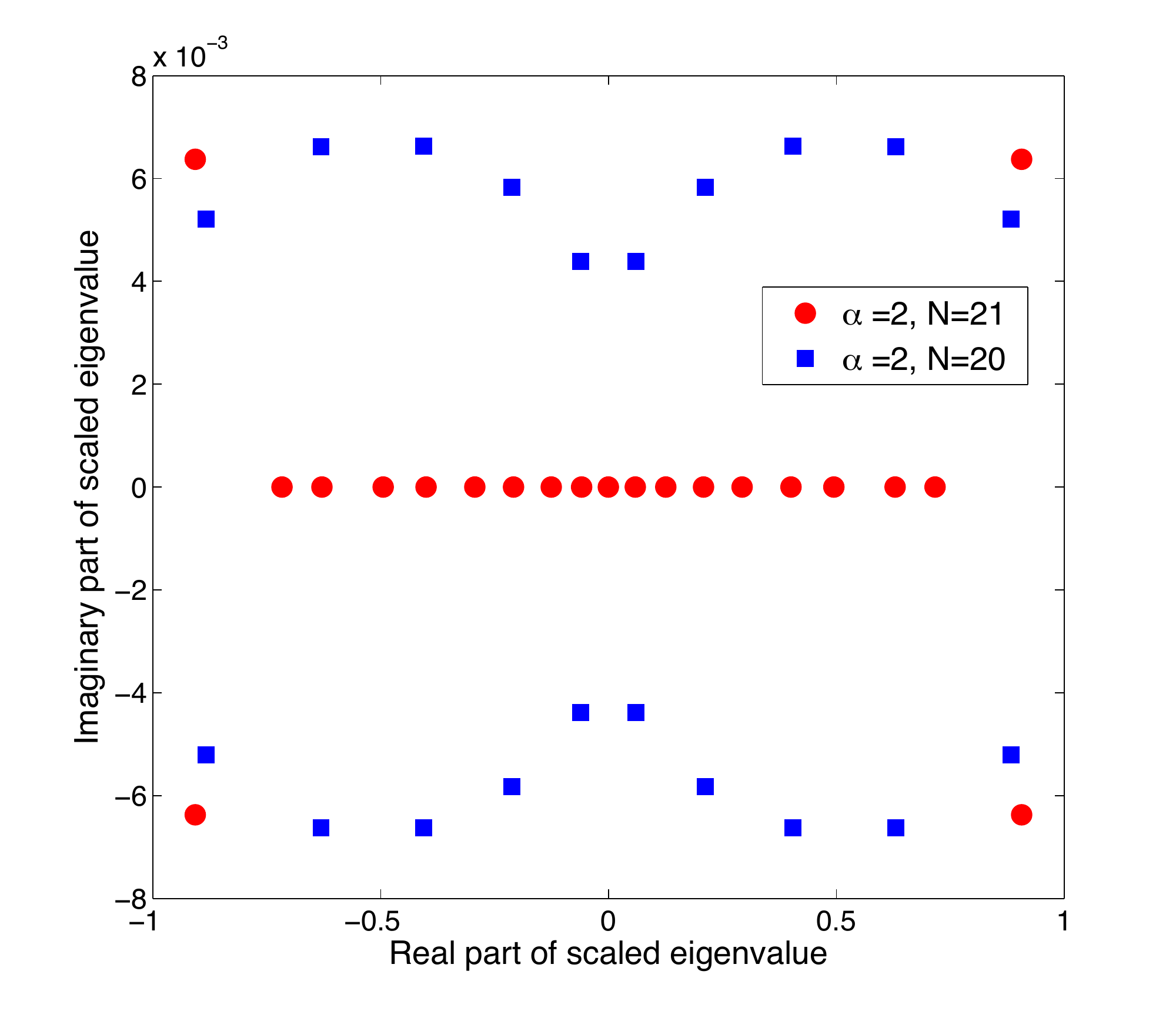}
\caption{(color online) Degrees of $\mP\mT$-symmetry breaking for an $\alpha=2$ lattice with closest impurities in the odd ($N=21$, red circles) and even ($N=20$, blue squares) cases. The complex eigenvalues are scaled by half-bandwidth $2\Delta_\alpha$, and are obtained for impurity strengths just above their respective $\mP\mT$-symmetry breaking thresholds: $\gamma/\Delta_\alpha=0.63$ for $N=21$ and $\gamma/\Delta_\alpha=1.08$ for $N=20$. When $N$ is odd (red circles), we see that four eigenvalues, near the top and the bottom of the energy band, become complex while the remaining $(N-4)$ remain real; when $N$ is even (blue squares), all $N$ eigenvalues simultaneously develop finite imaginary parts. This contrast becomes large for $N\gg 1$.}
\label{fig:complex}
\end{center}
\vspace{-5mm}
\end{figure}
 Figure~\ref{fig:complex} shows this generic even-odd effect for the case of closest impurities in a lattice with $\alpha=2$. For $N=20$ (blue squares), when $\gamma/\Delta_\alpha=1.08$, just above the even-lattice $\mP\mT$-symmetry breaking threshold, we see that all $N$ eigenvalues simultaneously become complex; on the other hand, for $N=21$ (red circles), when $\gamma/\Delta_\alpha=0.63$, just above the odd-lattice $\mP\mT$-symmetry breaking threshold, four eigenvalues near the band-edges become complex and the rest remain real. This difference has dramatic consequences.  


\noindent{\it Wavepacket Evolution:} We now explore the signatures of sequential, or simultaneous, $\mP\mT$ symmetry breaking in the time-evolution of a single-particle wavepacket. It is straightforward to obtain the site- and time-dependent intensity $|\langle k|\psi(t)|^2$ where $|\psi(t)\rangle= G(t)|\psi(0)\rangle$, $G(t)_{pq}=[\exp(-i\hat{H}_\alpha t/\hbar)]_{pq}$ is the time-evolution operator, $\hbar=h/(2\pi)$ is the (scaled) Planck's constant, and $\langle k|\psi(0)\rangle$ denotes the initial state. We remind the reader that since the Hamiltonian $H_\alpha$ is not Hermitian, $G(t)$ is not a unitary matrix. With an open (quantum) system in mind, we use the standard inner-product to obtain the time-dependent intensity; therefore, the maximum intensity at a site, for a normalized initial state, can exceed unity~\cite{longhi}. 

\begin{figure}[htb]
\begin{center}
\includegraphics[angle=0,width=10cm]{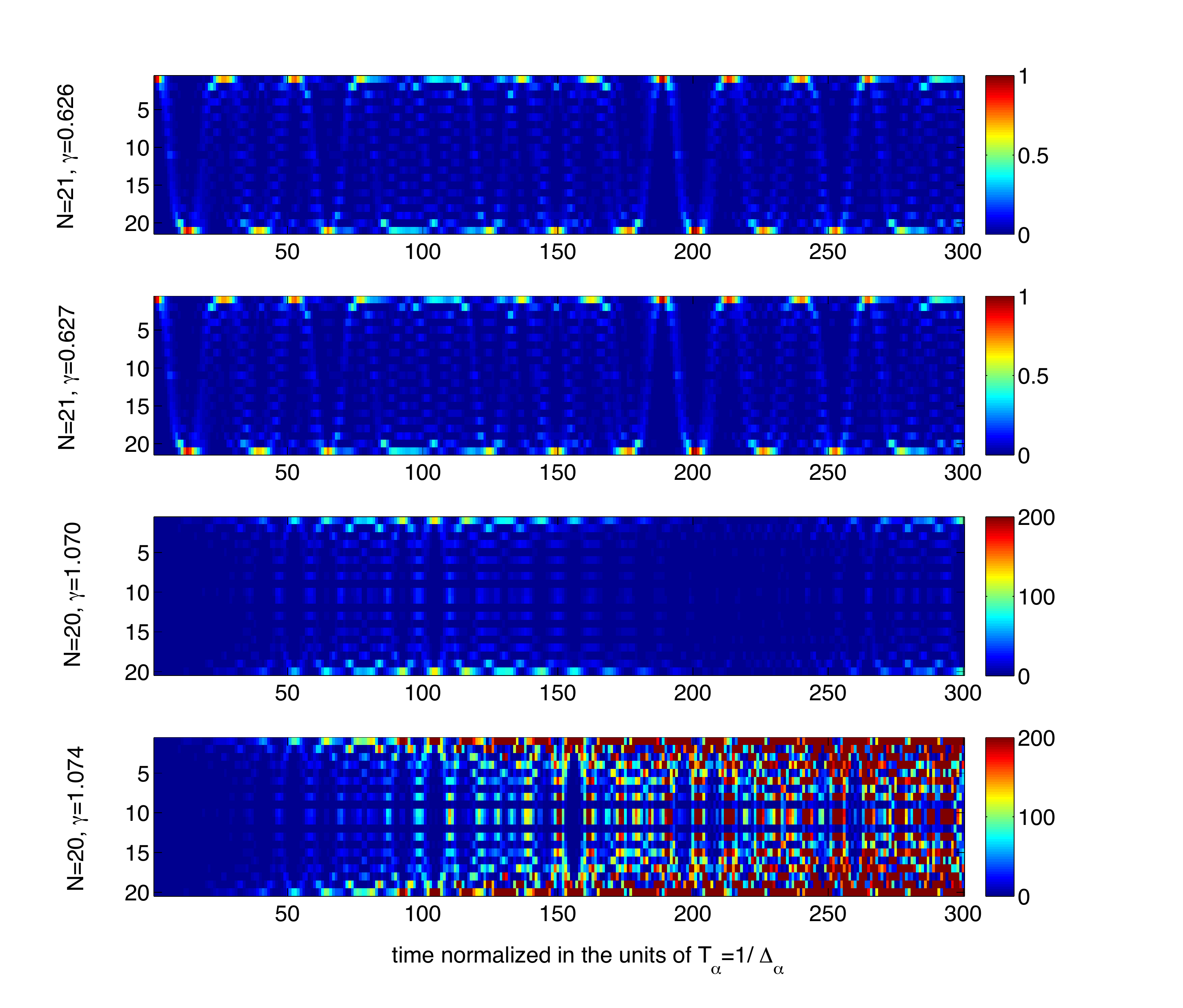}
\caption{(color online) Site- and time-dependent intensity profile for a particle initially localized on the first site in an $\alpha=2$ lattice. The vertical axis in each panel represents the site index. The top two panels are for an odd lattice, $N=21$, with impurities $\pm i\gamma$ at positions ($10,12$). They show only a small change in the intensity profile as $\gamma/\Delta_\alpha$ is increased, from 0.626 to 0.627, across the $\mP\mT$-symmetry breaking threshold and four eigenvalues become complex. The bottom two panels are for an even lattice, $N=20$, with nearest neighbor impurities at positions $(10,11)$. They show a dramatic change in the intensity profile that occurs when $\gamma/\Delta_\alpha$ is increased, from 1.070 to 1.074, across the $\mP\mT$-symmetry breaking threshold and all $N$ eigenvalues simultaneously become complex. This even-odd effect for $\mu\approx 1/2$ is independent of $N\gg 1$ or $\alpha$.}
\label{fig:dob}
\end{center}
\vspace{-5mm}
\end{figure}
Figure~\ref{fig:dob} shows the evolution of the site- and time-dependent intensity across the $\mP\mT$-symmetric phase boundary for a particle initially localized on the first site, $\langle k|\psi(0)\rangle=\delta_{k,1}$, in an $\alpha=2$ lattice. The vertical axis in each panel represents the site index. The top two panels show the predicted intensity profile for an odd lattice ($N=21$) with closest $\mP\mT$-symmetric impurities $\pm i\gamma$, at positions $(10,12)$. The first panel shows intensity oscillations that occur as the particle bounces from one end of the lattice to the other while encountering the (gain and loss) impurities with strength $\gamma/\Delta_\alpha=0.626$ at the center. When $\gamma/\Delta_\alpha=0.627$, the $\mP\mT$-symmetry is broken with four complex eigenvalues (Fig.~\ref{fig:complex}). We see that over the time-scale shown in Fig.~\ref{fig:dob}, the intensity distribution does not change significantly. The bottom two panels are for an $N=20$ lattice with impurities at positions $(10,11)$. The third panel shows the intensity oscillations that occur in the $\mP\mT$-symmetric phase when $\gamma/\Delta_\alpha=1.070$. When the impurity strength exceeds the critical value, $\gamma/\Delta_\alpha=0.174$, {\it all} $N$ eigenvalues become complex, thus maximally breaking the $\mP\mT$ symmetry (Fig.~\ref{fig:complex}). This property is manifest in the predicted intensity profile in the fourth, bottom-most panel. Note that we have scaled the color-bars in the bottom two panels to an identical range, to simplify comparison while maintaining the visibility of structures in the intensity profile; the maximum intensity in the bottom-most panel is $\sim 10^4$. 

\begin{figure}[htb]
\begin{center}
\includegraphics[angle=0,width=9cm]{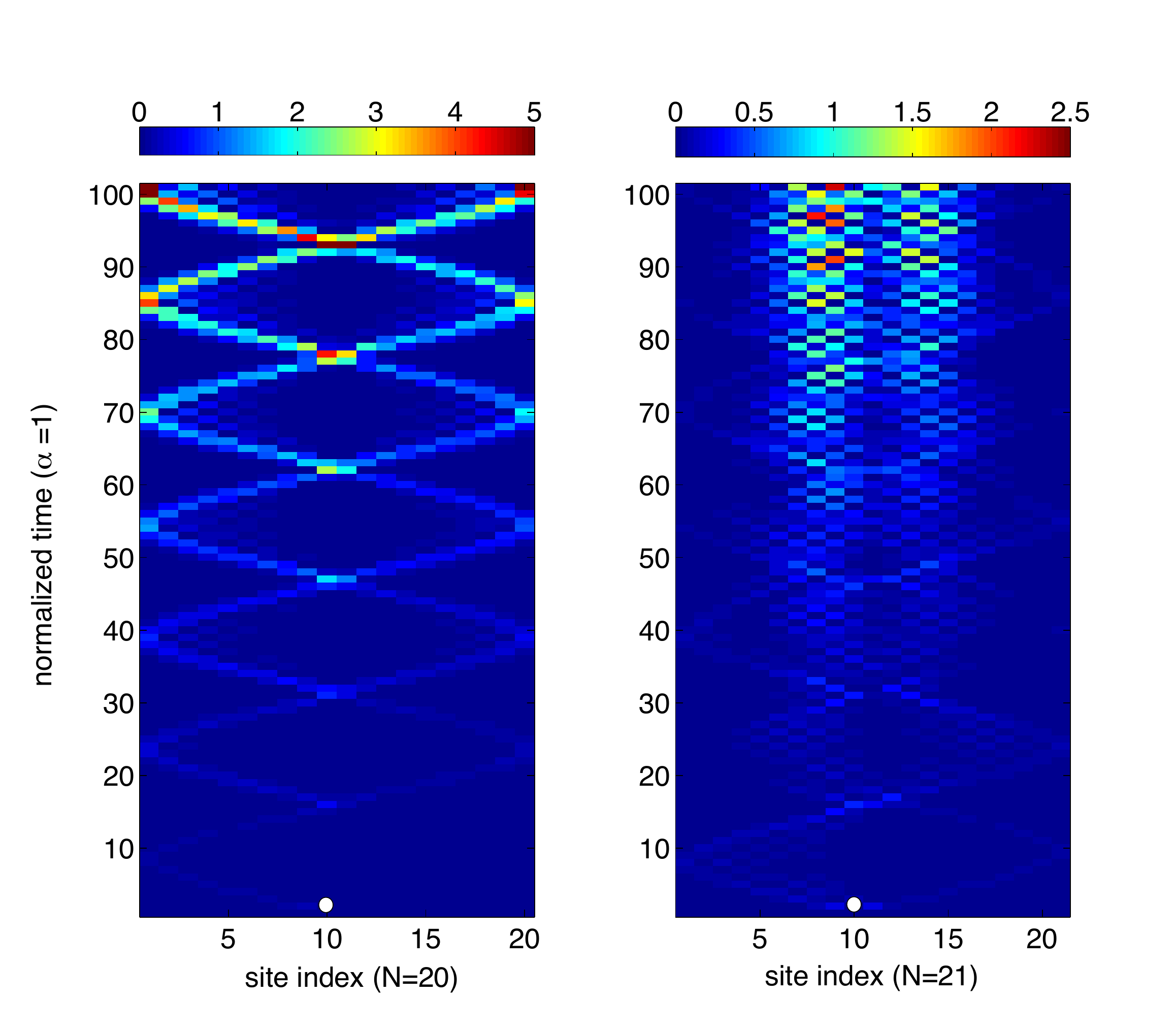}
\caption{(color online) Site- and time-dependent intensity profile for a state initially localized on the time-dependent-gain impurity, $m_0=10$, shown by a white dot, in an $\alpha=1$ lattice. The vertical axis denotes time normalized in units of $T_\alpha=1/\Delta_\alpha$, and the gain-impurity time-constant is $\tau/T_\alpha=5$. The intensity range for the even lattice (left panel) is larger than that for the odd lattice (right panel). The $\alpha=1$ lattice spectrum has equidistant eigenvalues~\cite{ya,longhi}, which gives rise to periodic behavior seen in both, even and odd, cases at times $t/T_\alpha\lesssim 50$. At large times, the $\mP\mT$-symmetry is maximally broken for the even, $N=20$, lattice, leading to commensurate real parts for the complex eigenvalues, and persistence of a "periodic" structure in the intensity; in contrast, for the odd, $N=21$, lattice, the commensuration partially is lost due to the emergence of only four complex eigenvalues. The even-odd effect does not depend upon the commensurate nature of the lattice spectrum and is valid for any $\alpha>0$, including a lattice with constant hopping, $\alpha=0$.}
\label{fig:time}
\end{center}
\vspace{-5mm}
\end{figure}
Lastly, motivated by the recent experimental setup~\cite{expt2}, we consider time evolution in a system where the impurities are given by $(\gamma_e(t),\gamma_L)$ at positions $(m_0,\bar{m}_0)$ respectively, where $\gamma_L>0$ is the loss and $\gamma_e(t)=\gamma_G(t)-\gamma_L$ is the effective, time-dependent, gain~\cite{expt2}. We choose $\gamma_G(t)=2\gamma_L(1-e^{-t/\tau})$ and $\gamma_L=\gamma_{PT}+0^{+}$, so that when $t/\tau\gg 1$, the system is (barely) in the broken $\mP\mT$-symmetric phase. Figure~\ref{fig:time} shows the remarkable even-odd effect that occurs due to different degrees of $\mP\mT$-symmetry breaking. The left (right) panel corresponds to results for an $\alpha=1$ lattice with $N=20,\gamma_L/\Delta_\alpha=1.06$ ($N=21,\gamma_L/\Delta_\alpha=0.60$),  and gain-impurity position $m_0=10$. The initial state $\langle k|\psi(0)\rangle=\delta_{km_0}$, localized at the gain site, is indicated by the white dot in both panels. The time-constant $\tau=5 T_\alpha=5/\Delta_\alpha$ is chosen so that the $\mP\mT$-symmetry condition, $\gamma_e(t)=\gamma_L$, is quickly achieved. We see that the overall intensity for the even lattice (left panel) is higher than that for the odd lattice (right panel), consistent with a higher number of complex eigenvalues. We emphasize that if the impurities are not the closest, the qualitative difference between time-evolutions in even and odd lattices is small; in both cases, then, the $\mP\mT$-symmetry breaks sequentially and the initial number of complex eigenvalues is the same. These results are valid for a lattice with any $\alpha$, including a constant hopping, and a generic initial state. {\it They predict that the degree of $\mP\mT$-symmetry breaking will be clearly visible in intensity measurements}, and that even and odd lattices will have dramatically different signatures of a broken $\mP\mT$-symmetry when $N\gg 1$.  

\noindent{\it Discussion:} In this paper, we have explored the robustness of the $\mP\mT$-symmetric phase, as well as the degree and signatures of its breaking in a lattice with a position-dependent, parity symmetric, hopping function $t_\alpha$ and a pair of imaginary, $\mP\mT$-symmetric impurities. We have shown that when the hopping is maximum near the center of the lattice, $\alpha>0$, the $\mP\mT$-symmetric phase lasts over an appreciable range of impurity strength, and that it is robust when the impurities are closest. The nature of $\mP\mT$-symmetry breaking transition at the robust point in even and odd lattices is different: when $N\gg 1$ is even, all eigenvalues simultaneously become complex, whereas when $N\gg 1$ is odd, only four of them initially do. We have predicted that this difference leads to clear signatures in the intensity profile of a single-particle wavepacket. 

In these calculations, we have ignored on-site disorder effects that, in coupled optical waveguides, arise due to variation in the real part of the refractive index. The effect of disorder on the critical impurity strength, as well as the interplay between the disorder, that tends to localize the particle to its initial position, and an imaginary potential, that tends to localize the particle at the impurity location, will deepen our understanding of these lattice models. 
    
Our results suggest that $\mP\mT$-symmetry breaking in a (non-uniform) lattice with $N\gg 1$ sites will be accompanied by remarkable phenomena, such as the even-odd effect predicted here, with no counterparts in the $N=2$ system~\cite{expt1,expt2}. 



\begin{thebibliography}{99}
\bibitem{note} The $\mP\mT$-symmetric phase is fragile everywhere when $\alpha<0$. Therefore, we only focus on the $\alpha\geq 0$ region.
\end{thebibliography}
\end{document}